\documentclass[showkeys,showpacs,prd]{revtex4}
\usepackage{amsmath}
\usepackage{amssymb}
\usepackage{graphicx}
\usepackage{color}

\begin{document}

\title{
Cosmological constant, supersymmetry, nonassociativity, and Big Numbers

}

\author{
Vladimir Dzhunushaliev$^{1,2}$
\footnote{Email: v.dzhunushaliev@gmail.com}
}
\affiliation{$^1$
Dept. Theor. and Nucl. Phys., KazNU, Almaty, 050040, Kazakhstan; \\
$^2$ IETP, Al-Farabi KazNU, Almaty, 050040, Kazakhstan \\
}

\begin{abstract}
The nonassociative generalization of supersymmetry is considered.
It is shown that the associator of four supersymmetry generators has the coefficient $\sim \hbar/ \ell_0^2$ where $\ell_0$ is some characteristic length. Two cases are considered: (a) $\ell_0^{-2}$ coincides with the cosmological constant; (b) $\ell_0$ is the classical radius of electron. It is also shown that the scaled constant is of the order of $10^{-120}$ for the first case and $10^{-30}$ for the second case. The possible manifestation and smallness of nonassociativity is discussed.
\end{abstract}


\maketitle

\section{Introduction}

The observed value of the cosmological constant is smaller than that predicted by an effective local quantum field theory by a factor of $10^{120}$. This discrepancy has been called ``the worst theoretical prediction in the history of physics'' \cite{Hobson,Smolin}. The great interest would be represented by the appearance of cosmological constant in any formula, even if obtained by some qualitative reasonings.

It is shown in Ref.~\cite{Dzhunushaliev:2013by} that the idea of supersymmetry can be extended with the inclusion of nonassociativity into supersymmetry. The main idea presented there is that an associator of four supersymmetrical quantities $Q_a, Q_{\dot a}, Q_b, Q_{\dot b}$ can be connected with the angular momentum operator.

In this Letter we want to consider the proportionality coefficient in this relation and, using some qualitative reasonings, to show that it may contain the factor which can be identified either with the cosmological constant or with the classical radius of electron.

\section{Nonassociative decomposition of the angular momentum operator}

In Ref. \cite{Dzhunushaliev:2013by}, a nonassociative generalization of a supersymmetry algebra with the supersymmetry generators $Q_a, Q_{\dot a}$ (here $a=1,2, \dot a = \dot 1, \dot 2$)  is considered. The simplest supersymmetry algebra considered there is (in this section we follow Ref.~\cite{Dzhunushaliev:2013by}):
\begin{eqnarray}
  \left\{
     Q_a , Q_{\dot a}
  \right\} &=& Q_a Q_{\dot a} + Q_{\dot a} Q_a =
  2 \sigma^\mu_{a \dot a} P_\mu,
\label{1-1}  \\
  \left\{
     Q_a , Q_b
  \right\} &=& \left\{
     Q_{\dot a} , Q_{\dot b}
  \right\} = 0,
\label{1-2} \\
  \left[ Q_a , P_\mu \right] &=& \left[ Q_{\dot a} , P_\mu \right] = 0,
\label{1-3}\\
  \left[ P_\mu , P_\nu \right] &=& 0.
\label{1-4}
\end{eqnarray}
The anticommutator \eqref{1-1} connects the momentum operator $P_\mu = -i \partial_\mu$ (here $\mu = 0,1,2,3$) and the generators $Q_{a, \dot a}$. The Pauli matrices $\sigma^\mu_{a \dot a}, \sigma_\mu^{a \dot a}$ are
\begin{eqnarray}
  \sigma^\mu_{a \dot a} &=& \left\{
    \left(
      \begin{array}{cc}
        1 & 0 \\
        0 & 1 \\
      \end{array}
    \right),
    \left(
    \begin{array}{cc}
        0 & 1 \\
        1 & 0 \\
      \end{array}
    \right),
    \left(
    \begin{array}{cc}
        0 & -i \\
        i & 0 \\
      \end{array}
    \right),
    \left(
    \begin{array}{cc}
        1 & 0 \\
        0 & -1 \\
      \end{array}
    \right)
  \right\},
\label{1-20}\\
  \sigma_\mu^{a \dot a} &=& \left\{
    \left(
      \begin{array}{cc}
        1 & 0 \\
        0 & 1 \\
      \end{array}
    \right),
    \left(
    \begin{array}{cc}
        0 & 1 \\
        1 & 0 \\
      \end{array}
    \right),
    \left(
    \begin{array}{cc}
        0 & i \\
        -i & 0 \\
      \end{array}
    \right),
    \left(
    \begin{array}{cc}
        1 & 0 \\
        0 & -1 \\
      \end{array}
    \right)
  \right\}.
\label{1-30}
\end{eqnarray}
Let us define an associator as follows:
\begin{equation}\label{1-70}
  \left[ x,y,z \right] = \left( x y \right) z - x \left( y z \right).
\end{equation}
It is assumed that the associator
$\left[ Q_a, Q_{\dot a}, \left( Q_b Q_{\dot b} \right) \right]$ is
\begin{equation}\label{1-80}
  \left[ Q_a, Q_{\dot a}, \left( Q_b Q_{\dot b} \right) \right] = 2 \zeta
  \sigma^\mu_{a \dot a} \sigma^\nu_{b \dot b} M_{\mu \nu},
\end{equation}
where the operator
\begin{equation}\label{1-90}
  M_{\mu \nu} = x_\mu P_\nu - x_\nu P_\mu
\end{equation}
is the angular momentum operator, and $\zeta$ is the still undefined numerical factor that equalizes the dimensions of the right- and left-hand sides of equation \eqref{1-80}. Our main goal here is to derive this factor using some plausible physical arguments.

\section{Definition of $\zeta$}

We see from equation \eqref{1-1} that the dimension of $Q$ is
\begin{equation}\label{2-10}
  \left[ Q \right] = \left(
    \frac{\text{g} \cdot \text{cm}}{\text{s}}
  \right)^{1/2} .
\end{equation}
From equation \eqref{1-80}, one can find that
\begin{equation}\label{2-20}
  \left[ \zeta \right] = \frac{\text{g}}{\text{s}}.
\end{equation}
We think that $\zeta$ should be constructed from fundamental constants. In this case one can consider following possibilities. The first one is that
\begin{equation}\label{2-30}
  \zeta \propto \frac{c^3}{G},
\end{equation}
where $c$ is the speed of light and $G$ is the gravitational constant. But we think that the relation \eqref{1-80} is a quantum relation, and in some sense it should be similar to the commutation relation
\begin{equation}\label{2-40}
  \left[ x, \hat p_x \right] = i \hbar .
\end{equation}
This means that the Planck constant should be included in $\zeta$. For example, it can be done as
\begin{equation}\label{2-55}
  \zeta \propto \frac{\hbar}{\ell_0^2},
\end{equation}
where $\ell_0$ is some characteristic length and should be constructed from physical constants. One can find following possibilities:
\begin{equation}\label{2-50}
  \ell_0 = \begin{cases}
    \frac{1}{\sqrt{\Lambda}},  &
    \Lambda \text{ is the cosmological constant,} \\
    r_0 = \frac{e^2}{m_e c^2}                        &
    \text{ is the classical radius of electron}
\end{cases}
\end{equation}
Choosing $\zeta$ in the form \eqref{2-50} we can rewrite \eqref{1-80} in the dimensionless form
\begin{equation}\label{2-60}
  \left[
    \tilde Q_a, \tilde Q_{\dot a},
    \left( \tilde Q_b \tilde Q_{\dot b} \right)
  \right] =
  2 \tilde \zeta
  \sigma^\mu_{a \dot a} \sigma^\nu_{b \dot b}
  \tilde M_{\mu \nu},
\end{equation}
where the quantities with tildes are dimensionless, and
\begin{equation}\label{2-70}
  \tilde \zeta =
  \begin{cases}
    l_{Pl}^2 \Lambda & \approx 10^{-120}; \\
    \frac{G m_e e^2}{c^3 \hbar^2} & \approx 10^{-30}
  \end{cases}
\end{equation}
where $l_{Pl} = \sqrt{\frac{\hbar G}{c^3}}$ is the Planck length; $e, m_e$ are the charge and mass of electron. Finally, write \eqref{1-80} in the form
\begin{equation}\label{2-80}
  \left[ Q_a, Q_{\dot a}, \left( Q_b Q_{\dot b} \right) \right] =
  \begin{cases}
    2 \zeta_0 \hbar \Lambda
    \sigma^\mu_{a \dot a} \sigma^\nu_{b \dot b} M_{\mu \nu}; \\
    2 \zeta_0 \frac{\hbar}{r_0^2}
    \sigma^\mu_{a \dot a} \sigma^\nu_{b \dot b} M_{\mu \nu}
  \end{cases}
\end{equation}
where the coefficient $\zeta$ from equation \eqref{1-80} is defined as
$\zeta = \zeta_0 \tilde \zeta$ and $\zeta_0 = \pm 1, \pm i$ is now a dimensionless number. The inverse relation is
\begin{equation}\label{3-150}
  M_{\mu \nu} =
  \begin{cases}
    \frac{1}{8 \zeta_0} \frac{1}{\hbar \Lambda}
  \sigma_\mu^{a \dot a} \sigma_\nu^{b \dot b}
  \left[ Q_a, Q_{\dot a}, \left( Q_b Q_{\dot b} \right) \right]; \\
    \frac{1}{8 \zeta_0} \frac{r_0^2}{\hbar}
  \sigma_\mu^{a \dot a} \sigma_\nu^{b \dot b}
  \left[ Q_a, Q_{\dot a}, \left( Q_b Q_{\dot b} \right) \right]
  \end{cases}
\end{equation}
We see that there are different possibilities for choosing $\zeta$. Probably it may be due to the fact that the coefficient $\zeta$ in the relation \eqref{1-80} can be different for different physical situations. For example, on the large scales ($\sim$ Universe) $\ell_0 \approx 1/\sqrt{\Lambda}$ but on the micro scales ($\sim r_0$) $\ell_0 \approx r_0$.

\section{Discussion and conclusions}

We have shown that the nonassociative generalization of supersymmetry has some coefficient that can be associated with some characteristic length. After scaling this dimensionless factor shows how small can be the manifestation of possible nonassociativity in physics. For the first case $\zeta \sim c^3/G$ and the dimensionless $\tilde \zeta \sim 1$ that is too much. For the second case the scaled value of this constant is the product of the squared Planck length and the cosmological constant
$\zeta \sim \hbar \Lambda$, and consequently is $\approx 10^{-120}$. The
possible manifestation of nonassociativity in this case become apparent in quantum gravity on large scales since $\tilde \zeta$ contains $\hbar, G$ and $\Lambda$. For the third case the scaled value of this constant is
$\zeta \sim G m_e e^2/(c^3 \hbar^2)$, and consequently is
$\approx 10^{-30}$. The possible manifestation of nonassociativity in this case become apparent in quantum gravity + electrodynamics on micro scales since $\tilde \zeta$ contains $\hbar, G, e$ and $m_e$.

\section*{Acknowledgements}

This work was supported by the Volkswagen Stiftung and by a grant in fundamental research in natural sciences by the Ministry of Education and Science of Kazakhstan. I am very grateful to V. Folomeev for fruitful discussions and comments.

\end{document}